# Anelastic relaxor behavior of $Pb(Mg_{1/3}Nb_{2/3})O_3$

Hana Uršič[1] and Dragan Damjanovic[2,*]

[1]*Electronic Ceramics Department, Jožef Stefan Institute, Jamova cesta 39,*

*1000 Ljubljana, Slovenia*

[2]*Ceramics Laboratory, Swiss Federal Institute of Technology in Lausanne - EPFL,*

*1015 Lausanne, Switzerland*

Elastic storage modulus and loss of relaxor lead magnesium niobate ceramics, $Pb(Mg_{1/3}Nb_{2/3})O_3$, have been measured with dynamic mechanical analyzer in single cantilever mode in the temperature range from 170 K to 320 K and at frequencies from 0.1 Hz to 50 Hz. The dependence of the elastic susceptibility (inverse modulus) on temperature and frequency of the driving force has characteristics of typical relaxor behavior that can be well described with the Vogel-Fulcher law. The parameters of the Vogel-Fulcher relation exhibit similar values for the dielectric and anelastic relaxations. Similarities and differences between anelastic and dielectric relaxor behaviors are identified.

* Author to whom correspondence should be addressed. Electronic mail: dragan.damjanovic@epfl.ch.

Lead magnesium niobate, $Pb(Mg_{1/3}Nb_{2/3})O_3$ or PMN, is a typical and the most studied dielectric relaxor material.[1] As shown in Fig. 1, the dielectric permittivity $\varepsilon$ of PMN exhibits a broad peak at a temperature $T_m$. Around and below this temperature the $\varepsilon$ depends on frequency $\omega$ of the probing electric field. The origin of the relaxation (or relaxor behavior) in $\varepsilon(\omega,T)$ is usually attributed to the dynamics of polar nanoregions (PNR), which form inside nonpolar matrix below Burns temperature (about 600 K for PMN).[2] The PNR increase in size during cooling and their response to the electric field slows until it freezes at a temperature $T_f$. The relation between $\omega$, $T_m$ and $T_f$ can be described by Vogel-Fulcher equation, $\omega = \omega_0 \exp[-E_a / k_B(T_m - T_f)]$ where $k_B$ is the Boltzman constant, $E_a$ can be interpreted as an activation energy, and $\omega_0$ is a characteristic frequency.[3]

Ambiguity in interpretation of Vogel-Fulcher equation has been discussed by Tagantsev and Glazounov and an alternative formulation has been proposed.[4,5] Other insights into the nature of relaxors have been provided recently by first principle calculations.[6,7] Takenaka *et al.* thus propose a model of relaxors as a homogeneous random network of anisotropically coupled dipoles.[7] The Burns temperature, first identified in PMN by optical measurements,[2] has been recently reassessed following the diffuse scattering experiments with cold neutrons.[8] Additional characteristic temperatures have been identified in relaxor materials.[9] Clearly, the physics of relaxors is still not completely understood.

While most studies of properties of dielectric relaxors focused on dielectric behavior, dynamic mechanical response has been less investigated.[10-13] One would *a priori* expect that PNR in dielectric relaxors respond to dynamic mechanical fields in a similar fashion to what was observed in so-called "strain glass" of metal alloys;[14] that is, mechanical compliance of PMN should show a similar temperature-frequency behavior as dielectric permittivity but this has not been fully demonstrated so far. In fact, the mechanical and dielectric susceptibilities in PMN were reported to exhibit important differences in their temperature behavior. First, the maximum in elastic susceptibility for a given frequency is several times weaker than that in dielectric permittivity, while the peak appears broader for elastic than for dielectric susceptibility[11,13]. Importantly, because most studies of mechanical properties have covered a limited frequency range, the Vogel-Fulcher relationship has not been demonstrated in dielectric relaxors for mechanical properties. Carpenter *et al.*[13] showed that in their experimental studies of anelastic response of PMN the slope of $ln\tau$ versus $T$ (where $\tau = \tau_0 \exp[E_a / k_B(T - T_f)]$ is the relaxation time at $T = T_m$ and $\tau_0$ is inverse of the attempt frequency) measured at four frequencies leads to a slope which is significantly lower than the one observed in the dielectric permittivity. The authors suggested that the difference arises because mechanical and electrical fields probe different aspects of PNR dynamics. For example, 180° flipping of PNR (as proposed by Cross for superparaelectric model of relaxors[15]) would not be seen in measurements of elastic modulus and could thus explain at least some of the difference in behavior of mechanical and dielectric susceptibilities. Comparing data on anelastic and dielectric relaxation may thus help uncovering details of the physics behind relaxor behavior.

In this paper we demonstrate experimentally Vogel-Fulcher relationship in elastic properties of PMN ceramics measured over two and half orders of magnitude in frequency. It is shown that Vogel-Fulcher relationships with similar values of the parameters describe well both the dielectric and anelastic susceptibilities. The experiments confirm large difference in the strength of anelastic and dielectric relaxations and indicate some other differences in the temperature dependence of the two relaxations.

PMN ceramics were synthesized using PbO (99.9% purity), MgO (98%,), and $Nb_2O_5$ (99.9%) powders. A mixture of PbO, MgO, and $Nb_2O_5$ in the molar ratio corresponding to the stoichiometric $Pb(Mg_{1/3}Nb_{2/3})O_3$ was high-energy milled in a planetary mill. The ceramics were prepared by pressing isostatically with 300 MPa and sintering in a double alumina vessel in PMN packing powder at 1200°C for 2 h. The heating and cooling rates were 2°C/min. The density of the ceramics measured by Archimedes' method is ≈96% of the theoretical density. The median grain diameter is $d_{50}$= 1.98 $\mu$m ± 1.05 $\mu$m. The weight loss on sintering was not determined; however, the dielectric loss tangent measured at 300°C and 100 Hz is below 0.01 (not shown) indicating a low concentration of defects. After sintering the samples were cut, polished and annealed at 600°C. For the electrical measurements, Cr/Au electrodes were deposited by sputtering. The permittivity $\varepsilon = \varepsilon' - i\varepsilon''$ was calculated from the capacitance and phase angle data measured as a function of temperature with an L-C-R bridge. The mechanical storage modulus, $E'$, and loss, $E''$, were measured in the single cantilever mode with a Perkin-Elmer PYRIS Diamond Dynamic Mechanical Analyzer (DMA). Mechanical susceptibility was calculated by taking inverse of the complex modulus. It should be understood that single cantilever measurements, which are the best choice in DMA technique for materials with a high stiffness, may give rather large errors in the absolute value of the elastic modulus (errors of 50% are not uncommon) so that numerical values given here and in the literature are only indicative[16]. The relative trends (such as temperature or frequency dependence of the modulus) of interest here are, however, very reliable. Samples used for mechanical measurements were rectangular bars with thickness of about 0.6–1 mm, length of 25–30 mm and width of 3–4 mm. Dielectric measurements (see Figure 1) were made on smaller samples cut form such a bar, with dimensions roughly 3x4x1 $mm^3$. Cr/Au electrodes were sputtered on large surfaces of those samples. The dielectric and elastic measurements were made during cooling at a temperature rate of about 2°K/minute.

Elastic susceptibility $(1/E)'$ (the real part of inverse of the elastic modulus $E$) and loss $(1/E)''$ for PMN ceramics are shown in Figure 2. We choose to show $1/E$ instead of $E$ for an easier comparison with the electric permittivity, as both indicate susceptibility of the material to respective external fields. The relaxor nature of the anelastic response is obvious from Figure 2. As reported by other authors who carried out measurements taken at a single or a few frequencies only, the dielectric peak is much stronger but narrower than elastic, Figure 3. This feature is here demonstrated over a frequency range covering two and a half orders of magnitude, but only one frequency is shown in Figure 3 to avoid data clutter.

Figure 4 plots $T_m$ versus $\ln(\omega)$ for both dielectric data from Figure 1 and elastic data from Figure 2. The fits with Vogel-Fulcher relations are shown as full lines. The agreement between dielectric and anelastic data is excellent, notwithstanding a slightly lower $T_m$ values for the anelastic data. The discrepancy is only 2–3 K and this can be easily accounted for by different positions of the samples with respect to the thermocouple in the two experimental set-ups. As indicated in the legend of Figure 4, the agreement among the Vogel-Fulcher parameters obtained separately for the dielectric and anelastic data is very good and is also in a good agreement with values obtained previously for dielectric relaxation in other studies.[5,17]

Comparing now dielectric and anelastic data for a same frequency, Figure 3, several observations can be made. As reported by other authors[11,13] our data show that, compared to the respective background susceptibilities, the dielectric relaxation is much stronger than the elastic: over the covered temperature range the electrical permittivity changes by more than five times while the change in elastic susceptibility is less than 10% (compare Figures 1 and 2 and see summary in Figure 3). Cordero *et al.* report 40% increase in elastic susceptibility of PMN modified with 10% $PbTiO_3$[11], while data for PMN obtained at high frequencies all show changes below 20%.[13] The weaker anelastic relaxation could, of course, be just a consequence of a large background elastic susceptibility that does not participate in the relaxation process. We assume here that nonlinear contributions may be excluded from consideration and that amplitudes of both the driving electric field (1 V) and elastic force (<1 N) applied on the samples could be considered as a weak field regime, so that it is justified to compare the elastic and dielectric spectra.

Another reason for the difference in the relaxation strength could be that two or more aspects of PNR dynamics contribute separately to the elastic and dielectric susceptibilities. An example would be 180° flipping of PNR which (if, in fact, present at all in PMN[15,18]) would contribute only to the dielectric permittivity but not elastic susceptibility. However, to the extent that Vogel-Fulcher relation accurately describes the dynamics of PNR, the good agreement between Vogel-Fulcher parameters for the anelastic and dielectric relaxation shown here suggests that both relaxations have the same origin. If so, it would appear that electro-elastic response of PNR is simply more sensitive to the excitation by the electric than the mechanical stimulus. It is important to note here that a relatively weak anelastic relaxation strength seems to be characteristic of the elastic response in relaxor-like systems in general. For example, in the so-called metallic "strain glass,"[14,19] which exhibits anelastic relaxation qualitatively similar to that shown here for PMN, the storage modulus changes over the relaxation maximum region by about 20%.

Another notable feature of the anelastic and dielectric relaxations is that above $T_m$ the temperature dependences of permittivity and elastic susceptibility are qualitatively similar (see Figure 3) while below $T_m$, where frequency dispersion appears, the dielectric response hardens steeply with decreasing temperature whereas the elastic susceptibility decreases more gently and almost linearly. This might suggests reduced electro-mechanical coupling below $T_m$ in PMN and it will be interesting to see if such reduced coupling can be seen in other experiments (e.g., electric field induced piezoelectric response[20] or electrostriction) and whether theoretical models can account for such behavior.

In conclusion, it is shown that PMN exhibits true anelastic relaxor behavior, with parameters of the Vogel-Fulcher equation similar to those for dielectric relaxation. These common features as well as number of differences in anelastic and dielectric relaxations revealed in the present study and in earlier experiments present challenges that should be addressed in models interpreting relaxor behavior.

The authors acknowledge financial support of FNS PNR62 project No. 406240-126091 (DD) and SRA Programme P2-0105 and CoE NAMASTE (HU). Technical support by S. Drnovšek is gratefully acknowledged.

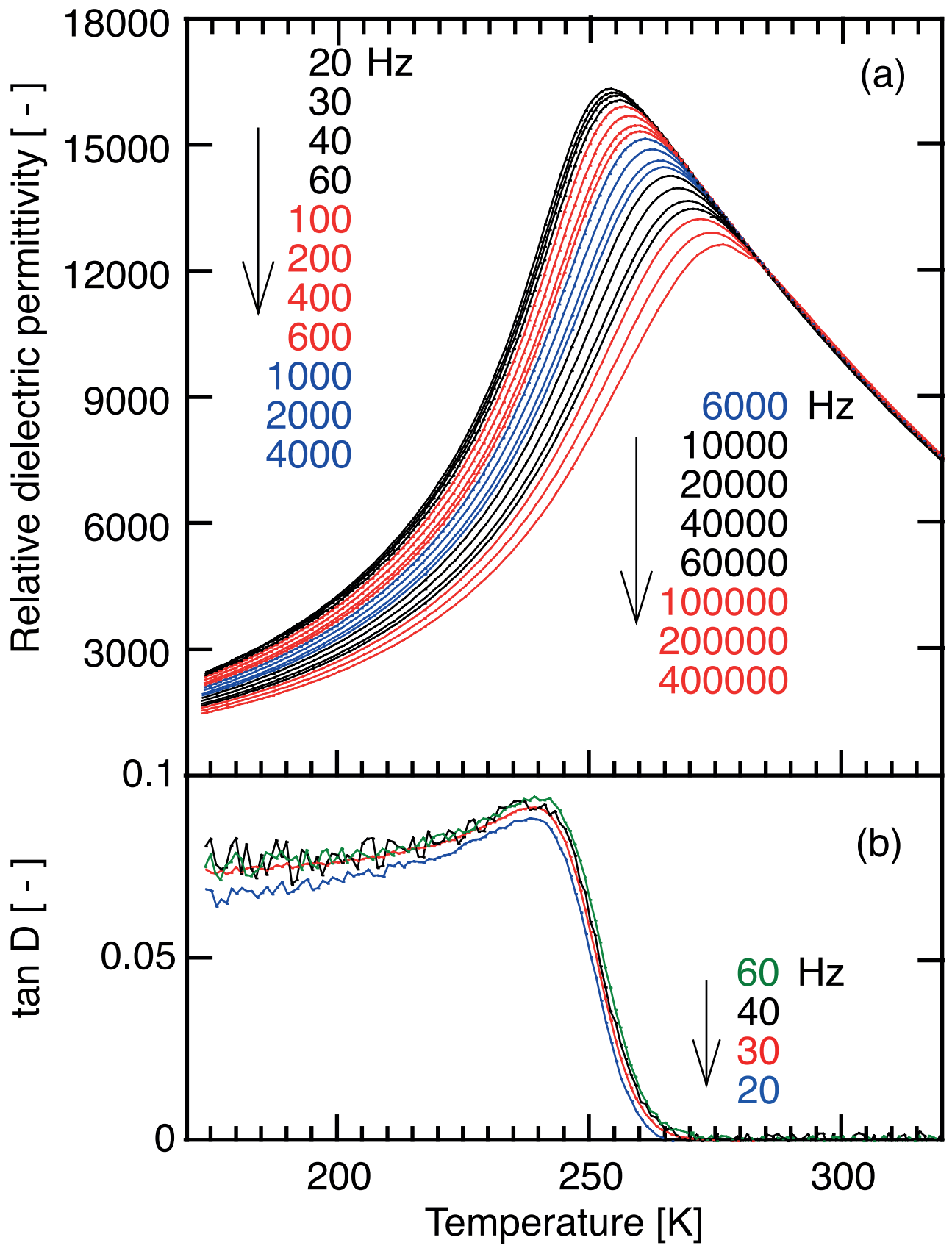

Figure 1. (a) The relative dielectric permittivity $\varepsilon' / \varepsilon_0$ ($\varepsilon_0$ is the electric constant) and (b) loss tangent ($\tan D = \varepsilon'' / \varepsilon'$) of PMN ceramics investigated in this study as a function of temperature, measured on cooling and at frequencies indicated in the figure.

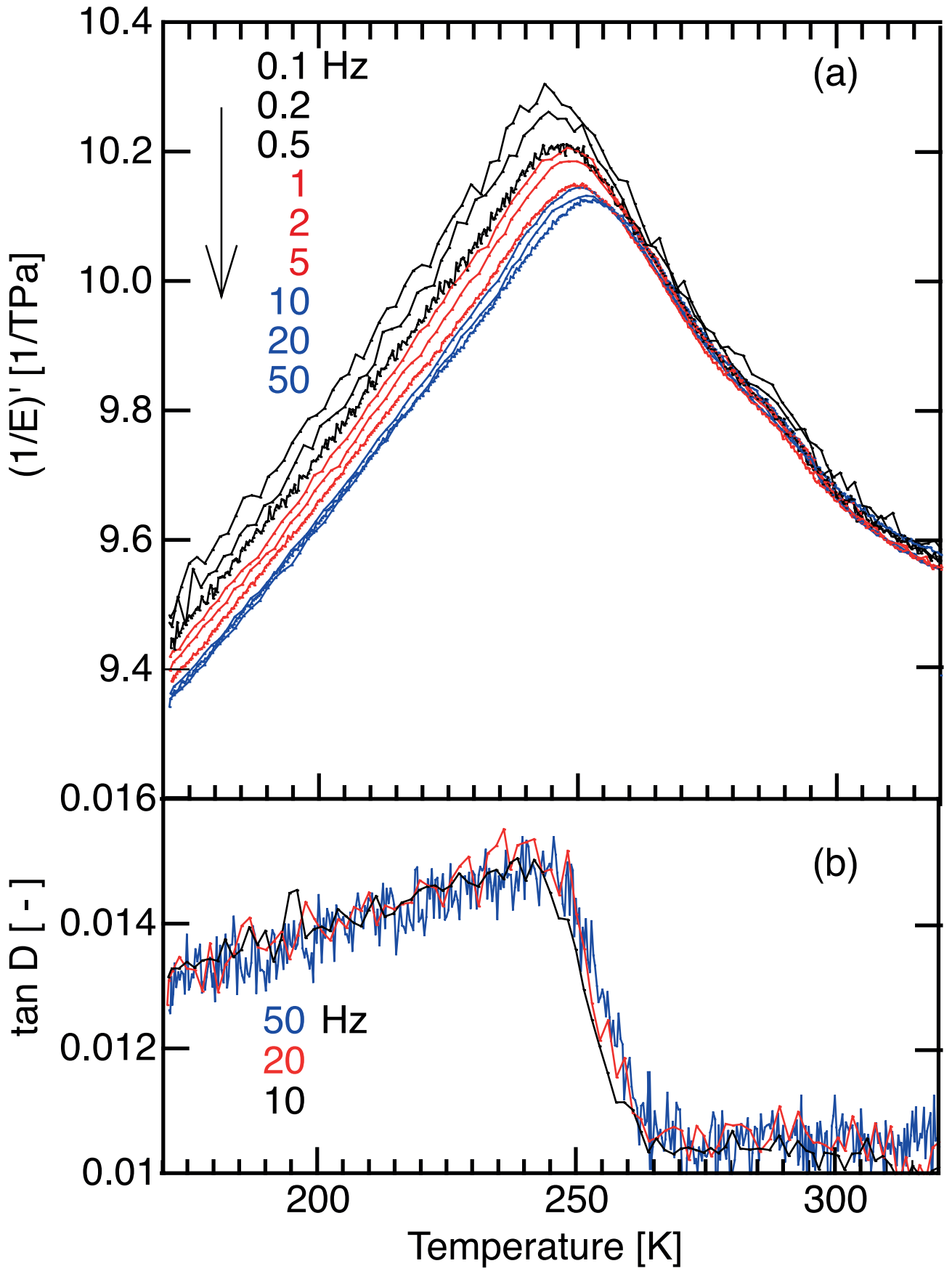

Figure 2. (a) Elastic susceptibility $(1/E)'$ (the real part of inverse of the elastic modulus $E$) and (b) loss tangent ($\tan D = E''/E'$) of PMN ceramics investigated in this study as a function of temperature, measured on cooling and at frequencies indicated in the figure. The numerical values for $(1/E)'$ are only indicative.

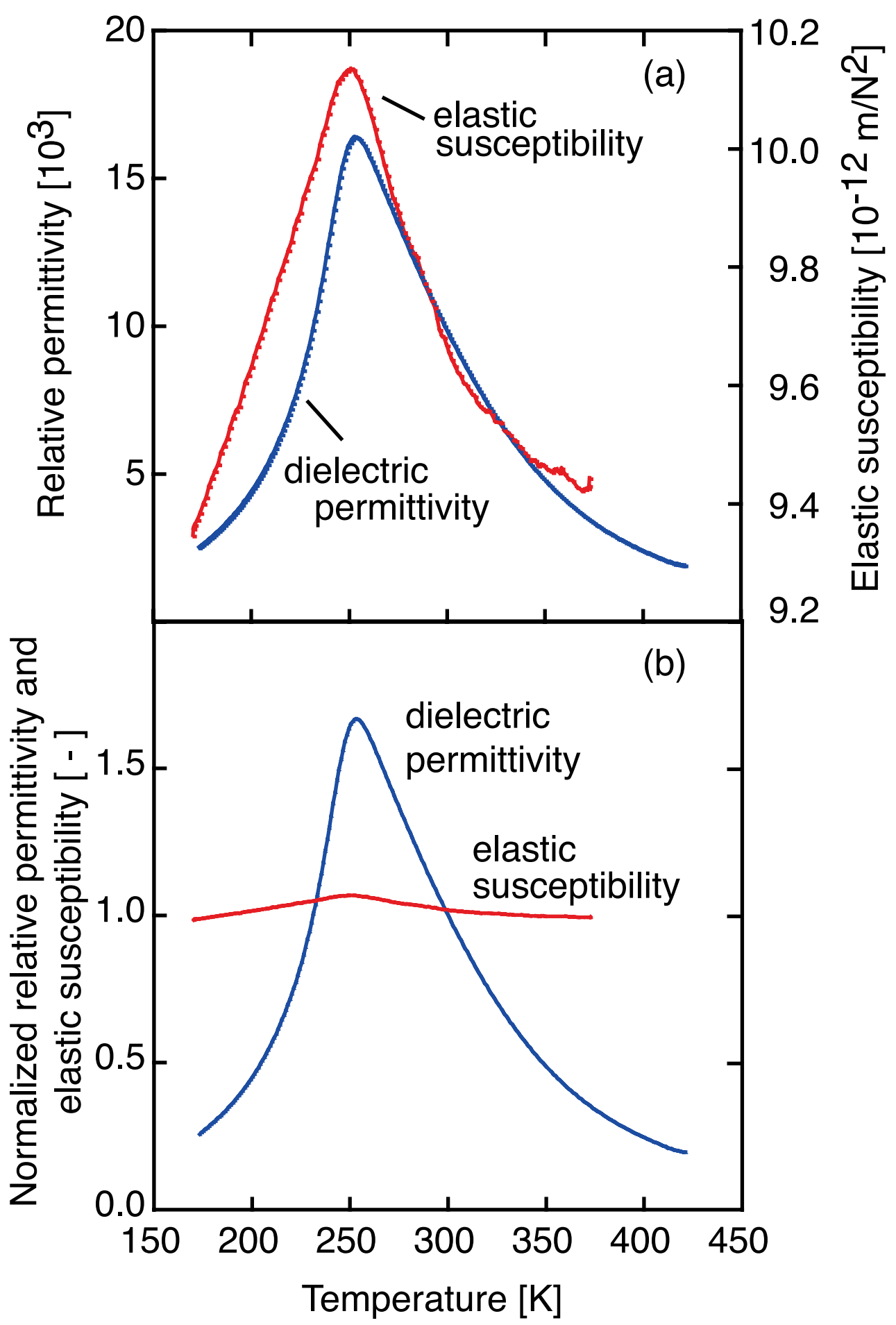

Figure 3. (a) Comparison of the relative dielectric permittivity and elastic susceptibility measured at 20 Hz. (b) The dielectric permittivity and elastic susceptibility from (a), normalized to the values at 300 K.

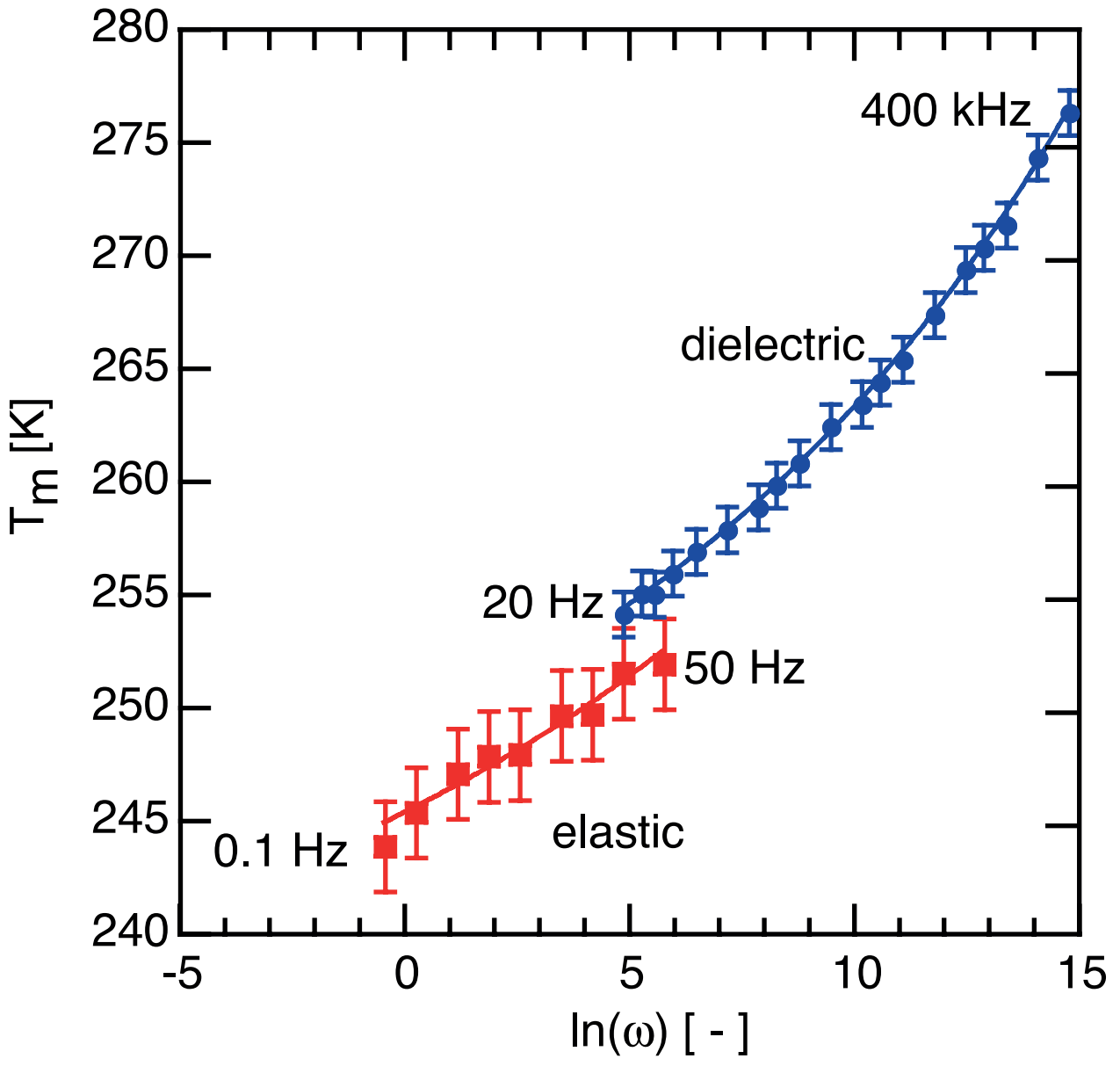

Figure 4. (a) The temperature $T_m$ of the maximum dielectric permittivity (circles) and elastic susceptibility (squares) from Figures 1 and 2, respectively, as a function of *ln*(ω). The full lines are fits through the data with Vogel–Fulcher equation (see text). The Volgel–Fulcher parameters obtained by fitting are $T_f \sim$ 217 K, E ~ 0.086 eV, $\omega_0 \sim 5 \times 10^{13}$ rad/s for the dielectric data, and $T_f \sim$ 215 K, E ~ 0.078 eV, $\omega_0 \sim 1.1 \times 10^{13}$ rad/s for the elastic data. If $T_m$ for the elastic data are increased by 2.5 K, the two $T_m$ – *ln*(ω) curves can be well fitted by a single Vogel-Fulcher equation with parameters very close to those obtained by fitting the dielectric data alone (not shown).